\documentclass[preprintnumbers,amsmath,amssymb]{revtex4}
\usepackage{graphicx}

\begin{document}
\title{Analysis of  White Dwarfs with Strange-Matter Cores}

\author{G. J. Mathews$^1$, I.-S. Suh$^2$, B. O'Gorman$^1$, N. Q. Lan$^1$, W. Zech$^1$, K. Otsuki$^3$, F. Weber$^4$}

\vskip .1 in
\affiliation{
$^1$Center for Astrophysics,
Department of Physics,
University of Notre Dame,
Notre Dame, IN 46556\\
$^2$Center for Research Computing,
University of Notre Dame,
Notre Dame, IN 46556\\
$^3$University of Chicago, Chicago,IL 60637\\
$^4$Department of Physics,
San Diego State University
San Diego, CA 92182}

\date{\today}
\begin{abstract}
We summarize masses and radii for a number of white dwarfs as deduced 
 from a combination of proper motion studies, {\it
Hipparcos} parallax distances, effective temperatures, and binary or
spectroscopic masses. A puzzling feature of these data, however, is
that some stars appear to have radii which are significantly smaller
than that expected for a  standard electron-degenerate white-dwarf
equations of state. We construct a projection of white-dwarf radii
for fixed effective  mass and conclude that there is at least
marginal evidence for bimodality in the radius distribution for
white dwarfs. We argue that if such compact white dwarfs exist it is
unlikely that they contain an iron core. We propose an alternative
of strange-quark matter within the white-dwarf core. We
also discuss the impact of the so-called color-flavor locked (CFL) state in
strange-matter core associated with color superconductivity. We show
that the data exhibit several features consistent with the expected
mass-radius relation of strange dwarfs. We identify eight
nearby white dwarfs which are possible candidates for strange matter
cores and suggest observational tests of this hypothesis.

\end{abstract}

\maketitle
\section{Introduction}
The possible existence of strange-matter stars has been speculated
upon for some time. Most of the work concerning their nature and
origin  has focused on neutron stars (cf.~\cite{Glendenning92}).  Indeed, there have been many suggestions of
possible observational evidence  
 of neutron stars which are
too compact or have cooled  too rapidly to be comprised of normal nuclear matter.
Although the first reports \cite{Drake02,Slane02} of strange-matter
stars seem ruled out, other good candidates remain \cite{Xu05,Majczyna05,Schaffner05,Weber05,Bagchi06}.

In this paper we are concerned with the possible existence of white dwarfs with strange-matter cores
as has  been proposed  by Glendenning et al. \cite{Glendenning95a,Glendenning95b} who coined the term {\it strange dwarfs}. Although, the
central density and temperature of white dwarfs are too low to allow
a spontaneous transition to strangelets (such as may occur in hot
proto-neutron stars \cite{Alcock86}), strange-matter
white dwarfs could gradually form \cite{Glendenning95a,Glendenning95b} during
the progenitor main-sequence by the accretion of a strange-matter
nugget. Such nuggets could exist either as a relic of the early
universe or as an ejected fragment from the merger/coalescence  of
strange-matter neutron stars. Once captured by a star,
strange-matter nuggets would gravitationally settle to the center
and begin to convert normal matter to strange matter. This would
eventually lead to the formation of an extended strange-quark-matter
core in the white dwarf remnant.

However unlikely this paradigm  may seem, we nevertheless think it
worthwhile to examine the evidence for the possible existence of a
population of such peculiar white dwarfs. The most distinguishing
characteristic of the existence of strange dwarfs is that they must
have a smaller radius. This unavoidable consequence simply follows
from the fact that strange-matter has more degrees of freedom and
can therefore be more compact than ordinary electron-degenerate
matter. In this paper we expand on a previous study \cite{Vartanyan04}
aimed at  identifying nearby candidate white dwarfs with strange matter cores.
We review the 
evidence for two populations of white dwarfs, one of which is
more compact.  We present an updated list of nearby stars 
with best determined masses and radii which are 
consistent with the existence of a strange-matter core.
We propose some observational tests of this hypothesis.

\subsection{Data}

The quality and quantity of observational data on the white-dwarf
mass-radius relation has been  improved in recent years due to the
accumulation of expanded proper motion surveys (e.g.~\cite{Bergeron01}) 
and  the availability of {\it Hipparcos}
parallax distances  for a number of white dwarfs. For example,
Provencal et al.~\cite{Provencal98} used {\it Hipparcos} data to deduce
luminosity radii for 10 white dwarfs in visual binaries or common
proper-motion systems as well as 11 field white dwarfs.  Complementary {\it HST} observations  have also been made, for
example  to better determine the spectroscopy for {\it Procyon B}
\cite{Provencal02} and the pulsation of {\it G226-29} \cite{Kepler00}. 
{\it Procyon B} at first appeared as a compact star in
\cite{Provencal98}. In \cite{Provencal02}, however, this
star is now confirmed  to lie on the normal white-dwarf mass-radius
relation.  This fact was missed in \cite{Vartanyan04}.
Nevertheless, several other stars in this sample still
appear to be compact.

A summary of our adopted masses and radii for 22 nearby white dwarfs
[from \cite{Provencal98,Provencal02}] is presented in Table 1. When
more than one method was used to determine masses, the astrometric
mass was taken to be better than the spectroscopic mass [except for
{\it Stein 2051B}  for which we adopt the spectroscopic mass as in
\cite{Provencal02}]. Lowest priority was given to gravitational
mass estimates as the inferred masses and radii assume a model for the white dwarf
structure.  (Indeed, there is no evidence of 2 populations in the  survey of 
\cite{Bergeron01} which we attribute to this systematic effect).
All radii are based upon the Stefan-Boltzmann
radius inferred from the observed luminosity and {\it Hipparcos}
distances or the gravitational redshift.  We note that since the {\it Hipparcos} mission,
critical assessments have been made \cite{VanLeeuwen05}  of the of
the data quality, and although some problems have been identified, the catalogue as published 
remains generally reliable within the quoted accuracies and can continue to
be adopted here.  We also note that 
radii quoted in table 5 of 
\cite{Provencal98} for two of
the stars (L481-60 and G154-B5B) are inconsistent with the measured
gravitational redshift given in table 4 of that paper.  The radii listed in table 1
have been corrected to be consistent with the observed gravitational redshift.

\begin{figure}
\includegraphics[width=3.5in]{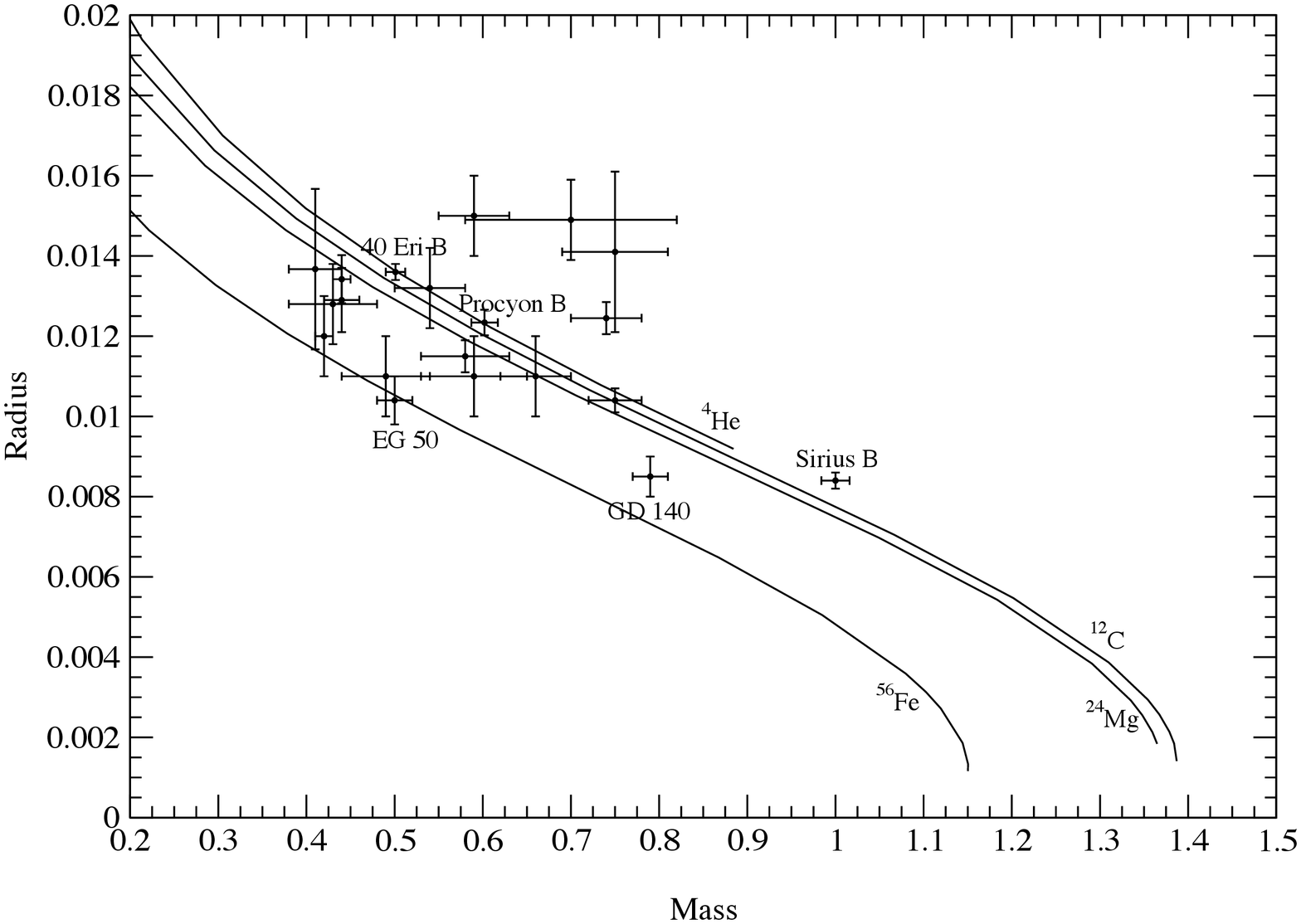} 
\caption[]{
The mass $M$ and radius $R$  for the 22 white dwarfs
\cite{Provencal98,Provencal02}.
The solid lines denote the Hamada \& Salpeter
model for normal  white dwarfs with the indicated composition.
}
\label{wdstandard}
\end{figure}

These data are compared with standard \cite{Hamada61}
mass-radius relations for He, C/O, and Mg white dwarfs in Figure
\ref{wdstandard}. It is readily apparent that some of the best
determined radii, e.g. {\it EG 50} and {\it GD 140}, seem
significantly more compact than that deduced from a normal
electron-degenerate equation of state.  {\it EG 50}  has a well
determined mass as it is a member of a visual binary. The mass of
the field white dwarf {\it GD 140} is also well determined.  It has
been well studied spectroscopically as its temperature is very high
so that this star is relatively bright. Nevertheless, both of these
stars are far more compact than expected for a normal white dwarf.
Indeed, in Ref.~\cite{Provencal98} it was concluded that an iron core might be
required to provide the required compactness as illustrated on
Figure \ref{wdstandard}.

On the other hand, other stars (e.g. {\it Sirius B}, {\it Procyon
B}, and {\it 40 Eri B}) fall nicely along the normal white-dwarf
mass-radius relation. Hence, there is a hint of evidence for the
existence of two white dwarf populations, one significantly more
compact than the other.

A straightforward projection of the distribution of white-dwarf
radii, however, obscures the two populations due to the fact that
one is also dealing with a distribution of masses.  To remove the
mass effect from the distribution of radii, we wish to construct
mass-radius relations which pass through each star. This is easily
facilitated by the fact that the mass-radius relations for different
compositions are nearly parallel for the range of masses of interest
here (cf. Fig.~1). Thus, curves which pass through each star can be
constructed from the parallel displacement of a single curve.

\begin{figure}
\includegraphics[width=4in]{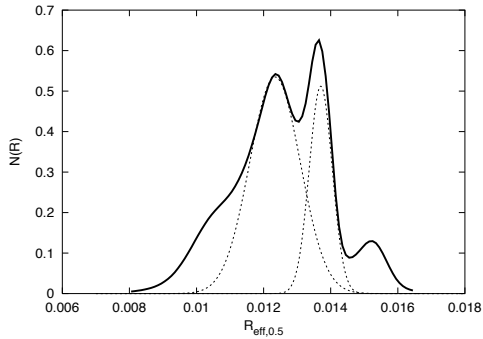} 
\caption[]{
Deduced distribution in radii for white dwarfs with a
 fixed effective $M = 0.5$ M$_\odot$.  Dotted lines
show gaussian fits to the two main peaks.
}
\label{wddist}
\end{figure}

The implied distribution of radii corresponding to a fixed
representative mass of 0.5 M$_\odot$ is shown on Figure
\ref{wddist}. This plot is constructed from the sum of Gaussian
distributions for each point with a width corresponding to the
uncertainty in the radius of each star. With only 22 stars in the
sample, the statistical and measurement errors are too marginal to
conclude that there is unambiguous evidence for a  bimodal
distribution.  Nevertheless, there is a hint of two peaks in this
distribution. One is a rather narrow peak corresponding to the
expected white dwarf radius around $R \approx  0.014 \pm 0.005$
$R_\odot$ corresponding to normal white dwarfs.  Below this peak
there is a somewhat broader distribution centered at $R = 0.012  \pm
0.010$ $R_\odot$. The total distribution is well fit with two
gaussians as shown by the dotted lines on Figure \ref{wddist}.

The lower end of the radius distribution is comprised of the eight
compact white dwarfs extending down to $R \sim 0.01$ $R_\odot$.
These are identified on the bottom of Table 1. These compact stars
may be members of  a compact population. In addition, above the
normal peak there are a few field stars with larger radii and large
uncertainties.  These presumably result from effects of the
white-dwarf model atmosphere used in the determination of the field
white dwarf radii \cite{Provencal98,Wood90} and are
probably not evidence for a population with large radii.

The  more-compact component of the radius distribution is roughly
what one might expect if a significant fraction of stars have
accreted a strange-quark nugget during their main-sequence lifetime.
Stars which have not developed a strange-matter core will be
narrowly centered around the normal white dwarf radius.  Those with
strange-matter interiors would be expected to have cores which will
have grown to their maximum size.  The corresponding radius is $\sim
80\%$ that of normal white dwarfs \cite{Glendenning92} as we now show.

\section{The Model}

To construct mass-radius relations we numerically integrate the
Tolman - Oppenheimer - Volkoff (TOV)  equation for general
relativistic hydrostatic equilibrium.
\begin{equation}
    {dP \over  dr}   =   - \biggl({ G M_r \over r^2}\biggr)
 { (\rho  + (P / c^2)) (1 + (4 \pi P r^3 / M_r c^2))  \over
   (1 - (2 G M_r / r c^2)) }~~.
\end{equation}
with $M_r$ the interior mass,
\begin{equation}
    M_r = \int_0^r 4 \pi r^2 \rho dr~~,
\end{equation}
and the pressure $P$ is related to the matter mass density $\rho$ through
an appropriate  equation of state (EOS).

For normal white dwarfs and normal matter in strange dwarfs we
utilize the standard  EOS  of \cite{Salpeter61,Hamada61}. This EOS  produces the mass-radius relationships shown in
Figure \ref{wdstandard}.  For these curves,  the central density,
$\rho_c$  varies from $10^5$ to $10^{10}$ g cm$^{-3}$. As noted
previously, a number of stars surveyed by {\it Hipparcos} have radii
which are less than the radius expected for a C/O white dwarf.

\section{Modifications of the WD EOS}
We have made a systematic study of all correction terms to the white
dwarf equation of state, e.g. Coulomb correction, lattice energy,
exchange energy, Thomas-Fermi correction, correlation energy,  etc.
We have even considered possible effects of magnetic fields \cite{Suh00}. 
None of these corrections can be reasonably varied to
fit these compact white dwarfs for a normal He,  C,  or even Mg
white dwarf.

\subsection{Iron-core white dwarfs}
As noted in \cite{Provencal98}, the only equation-of-state
parameter which can be modified to account for these data is the
composition.  The main dependence is through the charge to baryon
ratio $Z/A$ which appears in the dominant noninteracting
degenerate-electron term.
\begin{equation}
P_0 = {m c^2 \over 24 \pi^2} \biggl({m c \over \hbar}\biggr)^3 f(x)~~,
\end{equation}
where
\begin{equation}
f(x) = x (2 x^2 - 3) (x^2 + 1)^{1/2} + 3 \sinh^{-1}{x}~~,
\end{equation}
and
\begin{equation}
x = \biggl( {\hbar \over m_e c}\biggr)\biggl( {3 \pi^2 \over m_p} {Z \over A} \rho
\biggr)^{1/3}~~.
\end{equation}
In the nonrelativistic limit,
the white dwarf radius roughly scales as $(Z/A)^{5/3}$.
Decreasing  $Z/A$ from 0.5 to 0.46 appropriate to an iron composition
sufficiently diminishes  the radius so as to be consistent with the
most compact stars in this sample.

Achieving such an iron-core white dwarf, however, is difficult from
a stellar evolution standpoint. Single stars with $M \le 8$
M$_\odot$ are thought to terminate their evolution with the
formation of an electron-degenerate C/O core \cite{Iben83}.
Stars with $M \approx  8$  to 11 M$_\odot$ probably collapse during
nuclear statistical equilibrium burning without ever developing an
iron core \cite{Woosley86,Woosley95}. The formation of an iron
core only occurs during the final episode of quasi-equilibrium
silicon burning at the end of the evolution of a massive ($M \ge 11$
M$_\odot$) progenitor star. The final episode of silicon burning to
an iron core is exceedingly rapid both due to the high nuclear
burning temperatures and the fact that the energy content per gram
is diminished as the composition of the core shifts to heavier
nuclei. The formation of the iron core in silicon burning typically
lasts for only $\sim$ days and unavoidably proceeds until the core
mass exceeds the Chandrasekhar mass. The end result is collapse to a
proto-neutron star. Furthermore, even if one could somehow halt this
rapid thermonuclear burn to the Chandrasekhar mass, it is difficult
to imagine how to eject the outer layers of the star to expose the
inner iron-core white dwarf.

\subsection{Binary evolution}
Another possibility might be some sort of exotic binary evolution.
For example, Roche lobe overflow from one member of a binary can
expose the white-dwarf core during a common envelope phase \cite{Iben84,Iben85}. 
However, this process typically requires at
least $10^3$ to 10$^5$ years and is unlikely to be completed just at
the time at which the iron core is rapidly forming.

One could also imagine that a gradual mass deposition onto a white
dwarf might somehow lead to  episodic thermonuclear burning episodes
which could produce an iron core.  Effects of mass accretion onto a
white dwarf have been studied for some time 
(cf. \cite{Nomoto85,Woosley86,Iwamoto99}).
 Most mass accretion
rates ultimately lead to either a carbon detonation or deflagration
type Ia supernova which leaves no white-dwarf remnant. There is a
narrow region, however, for $M \sim 1.2$ M$_\odot$ and $\dot M \sim
10^{-9}$ in which helium detonates prior to carbon ignition so that
no carbon detonation occurs.  The resulting white-dwarf remnant,
however, does not contain an iron core. There does not seem to be a
middle ground.  The conditions necessary to burn white-dwarf
material to iron require such high densities and rapid reaction
rates that it would seem impossible to fine tune the parameters of
an accreting white dwarf to avoid the thermonuclear runaway
associated with a Type-Ia supernova and disruption of the star.

\section{Strange-Matter EOS}
Strange matter,  presents an alternative explanation for the
existence of compact white dwarfs. Strange stars have smaller radii.
This is simply because the density of strange quark matter is very
high ($> 2$ times nuclear matter density). The density is so high
for two reasons.  One is that the QCD vacuum energy does not
overwhelm the degeneracy pressure from the quarks and gluons until
high density.  Once the QCD vacuum energy dominates, one does not
require gravity to maintain strange matter. Strange-quark matter is
self bound and the star has no trouble to maintain the matter in
hydrostatic equilibrium. Another factor contributing to the high
density of strange-quark matter is the existence of three different
quark fermion species  within the core as opposed to only two in
ordinary quark matter, or only one in the case of a simple
electron-degenerate white dwarf.  The additional degrees of freedom
for strange-quark matter lower the degeneracy pressure and Fermi
energy and allow the matter to be more compact.

A strange white dwarf is expected \cite{Glendenning95a,Glendenning95b} to
consist of three distinct regions, a crust,  a core-crust boundary,
and a core.  Each region requires a different  EOS as we now
describe.

\subsection{Crust}
\label{Crust}

The crust of the star is composed of normal degenerate matter. For
the present purposes we take this crust to be predominantly composed
of $^{12}$C with a Hamada \& Salpeter (1961) equation of state.
Within the crust the density varies, but it is limited  to be less
than the neutron drip density, $\rho_{drip} = 4.3 \times  10^{11}$
g cm$^{-3}$.

\subsection{Crust-core boundary}
\label{boundary}
At densities higher than $\rho_{drip}$ free neutrons are released
from nuclei.  These would gravitate to the core where they would be
absorbed and converted into strange matter
 \cite{Glendenning92,Glendenning95a,Glendenning95b}.  As long as this
continues, the strange-matter core will grow. However, it is
expected \cite{Alcock86} that a sharp boundary
between the inner core and the outer crust will develop at the point
at which the crust density falls just below the neutron drip
density.

This gap between the strange-matter core and the outer crust
develops due to a Coulombic repulsion. Strange matter has a net
positive charge because the finite ($\sim 150$ MeV) mass of the
strange quark prevents them from forming in sufficient quantities to
maintain neutral strange matter.  The degenerate electrons are
unable to fully neutralize the positive charge of the inner
strange-matter core because they are not bound by the strong
interaction which keeps the core compact.  Hence, a net positive
charge exists in the strange-matter core. A dipole layer of very
high voltage develops between the crust and the core.. This high
voltage potential isolates the core from the outer crust which it
also polarizes.

A slight gap thus exists  between the core and crust regions. This
prevents the further growth of the core beyond the radius associated
with the neutron drip density. Hence, we take the radius at which
the neutron-drip density is achieved as the inner boundary of the
crust. This also defines the natural size to which the strange-quark
core can grow.

\subsection{Core}

\begin{figure}
\includegraphics[width=3in]{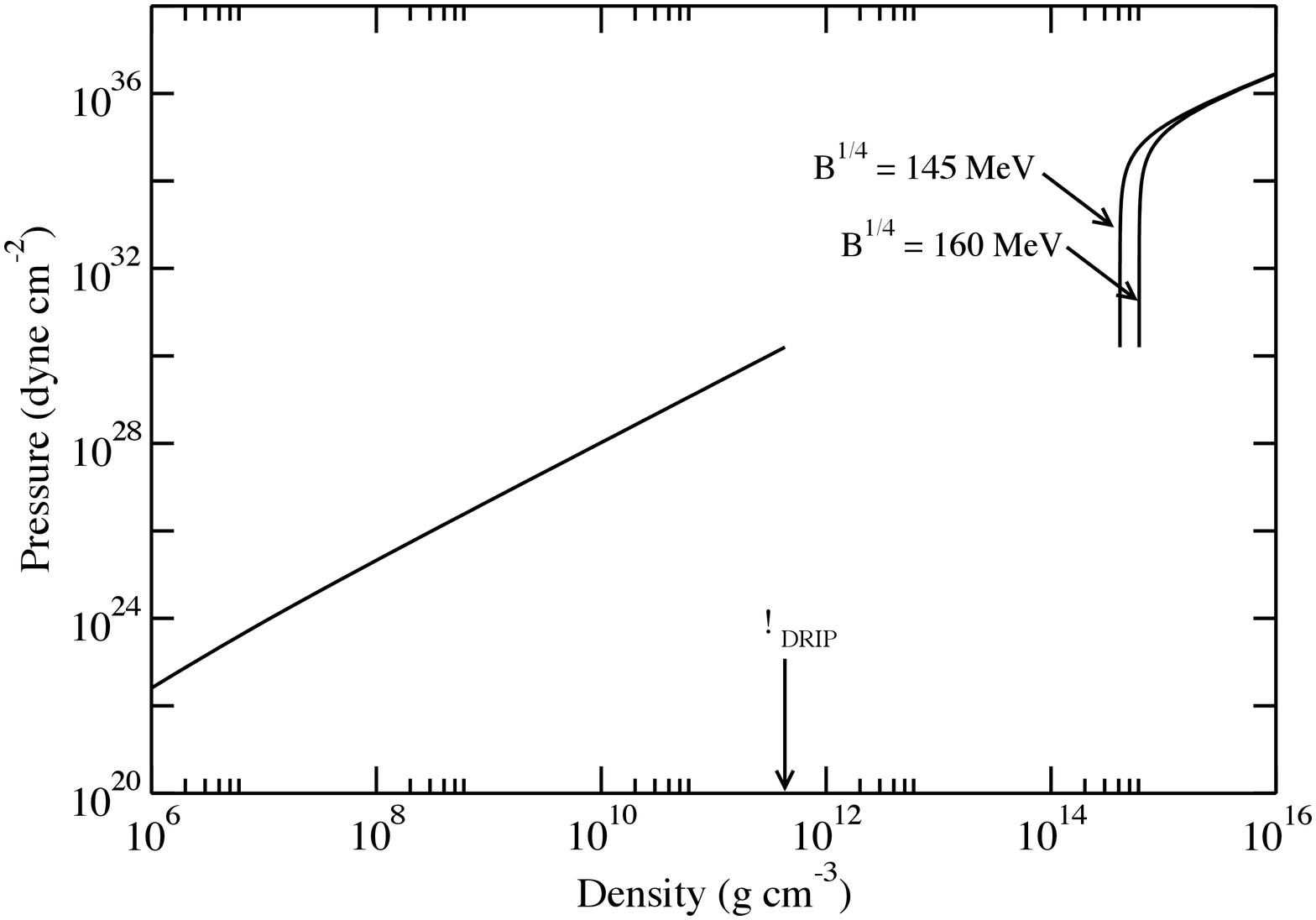} 
\caption[]{
Illustration of the strange-matter equation of state.
}
\label{wdeos}
\end{figure}

The core of the star is  taken to be comprised of up, down and
strange quarks.  The  {\it MIT} bag model equation of state is
adequate for our purposes. The pressure is thus
 related to density by
\begin{equation}
    P    =    (\rho  - 4 B)  /  3~~,
\label{seos}
\end{equation}
where $\rho$ is the mass-energy density contribution dominated by
noninteracting $\approx$ massless $u$ and $d$ quarks and gluons plus
an $s$ quark with $m \approx 150$ MeV.  The quantity  $B$ is the bag
constant which denotes the QCD vacuum energy. The bag constant is
constrained from hadronic properties to be, $B^{1/4}$ $\sim 145$ to
160 MeV. Within the quark core, the density is on the order of $2-3$
times nuclear matter density $(\sim 4-6 \times 10^{14}$ g cm$^{-3}$.
The equation of state for strange dwarfs is illustrated in Figure
\ref{wdeos}.  This shows the transition to strange quark matter
which takes place at the neutron drip density.  Although the density
steps discontinuously, the star remains in pressure equilibrium and
the pressure varies continuously through the star.

\begin{figure}
\includegraphics[width=3in]{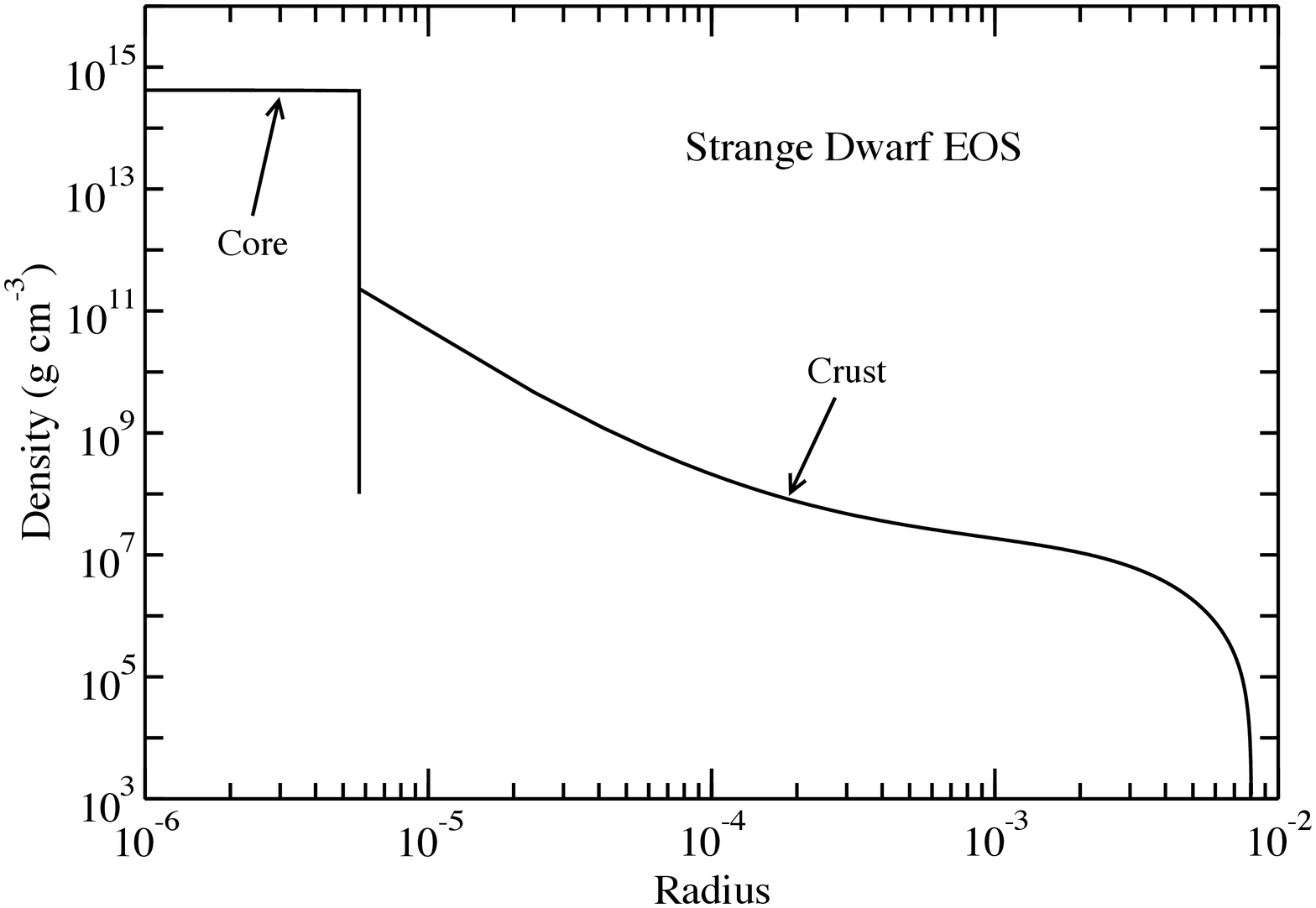} 
\includegraphics[width=3in]{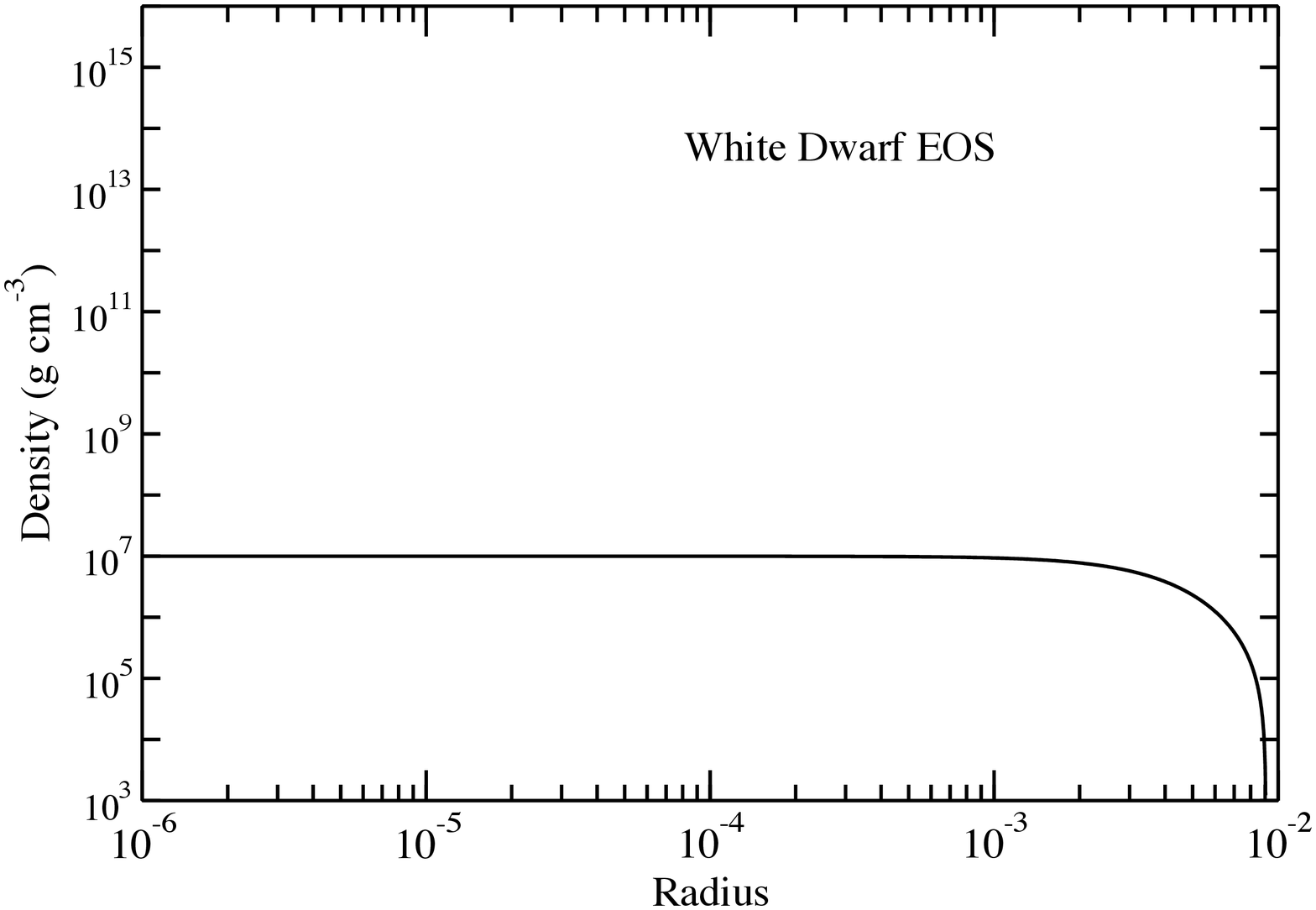} 
\caption[]{
Illustrations of the interior density vs. radius for a strange-matter white dwarf (upper figure)
and a normal-matter white dwarf (lower figure).
The separation between the core and crust regions is indicated in the upper curve.
}
\label{wdprofile}
\end{figure}

Having specified the EOS, we have radially integrated the TOV
equation for various initial central densities until the density
falls below a minimum density of $10^{-9} \times \rho_{drip}$.
Examples of the interior radial profiles of both a normal and a
strange-matter white dwarf are shown in Figure \ref{wdprofile}.
Here, one can see that most of the outer crust looks quite similar
to a normal white dwarf. A deviation from the normal white dwarf
density profile is only apparent for the inner few percent of the
radius of the star. Nevertheless, the existence of the compact inner
core leads to a smaller surface radius for the strange dwarf.

For models with $B^{1/4}  = 145$ MeV, the central density $\rho _c$
varies from 4.1 - $4.2 \times 10^{14}$ g cm$^{-3}$. In models with
$B^{1/4}  = 160$ MeV, $\rho_c$ varies from 6.1 - $ 6.2 \times
10^{14}$ g cm$^{-3}$. At these densities, strange dwarfs are in a
mass range comparable to that for ordinary white dwarfs, i.e. $0.3
M_\odot \le   M \le 1.35 M_\odot$. In \cite{Glendenning95a,Glendenning95b}  a
radial pulsation analysis is performed which demonstrates that these
strange dwarfs are indeed stable.

\subsection{Color Superconductivity}

Here a few remarks about the possible existence of color superconductivity in
strange-quark matter are warranted.  It is now generally 
accepted \cite{Alford99,Rajagopal01}
that quark matter at high
density may be in a so-called color-flavor locked (CFL) state, with
equal numbers of $u$, $d$, and $s$ quarks. This CFL state could
 be the ground state of strong interaction and therefore
stable even though the quark masses are unequal.  On the other hand,
quark matter at the "lower" densities of interest here \cite{Weber05} may be a 2-flavor
superconductor, 2SC, rather than a CFL superconductor. The 2SC state is
characterized by a much smaller pairing gap and
very similar to the simple bag model equation of state (Eq.~\ref{seos}) employed here.

Moreover, even for  CFL strange-matter, the thermodynamic potential is given as $\Omega_{CFL} =
\Omega_{free} - B_{eff}$, where $\Omega_{free}$ is the thermodynamic
potential for  ordinary unpaired quarks and $B_{eff} = - 3
\Delta^2 \mu^2 / \pi^2 + B$ \cite{Lugones03}, where
$\Delta$ is the gap of the $QCD$ Cooper pairs, $\mu$ is the average
quark chemical potential, and $B$ is the bag constant of Eq.~(\ref{seos}). 
In this case, $B_{eff}$ is
actually not constant but depends upon the chemical potential $\mu$.
However, by adopting  a characteristic mean value of the chemical potential
for CFL strange matter, we can set $B_{eff}$ as an overall constant.
With this simplification, if we can use this $B_{eff}$ instead of $B$ in
the Eq. (\ref{seos}), the only effect is an overall shift of structural properties by some
constant. Even for large gaps, the equation of state (Eq.~\ref{seos}) is only modified 
by a few percent of the bulk energy.  Such small effects can be safely neglected
in the discussions here.  Therefore, the equation of state of Eq. (\ref{seos}) remains 
adequate for our purpose by simply exchanging
  $B_{eff}$ for $B$.

Another issue associated with a CFL strange-matter core in
the configuration of strange dwarfs is halting  the growth of the
charge neutral CFL strange-matter core. In Ref.~\cite{Rajagopal01}
it has been shown that  the CFL phase is electrically
neutral in bulk without any need for electrons.  However,  as
mentioned in Sec. \ref{boundary}, in order to prevent the further growth of the
core, the strange-matter core should have a net positive charge.  This produces 
 a gap of very high voltage between the strange-matter core and
the outer crust  due to  Coulombic repulsion around the
neutron drip density. The CFL strange-matter consists
of equal numbers of $u$, $d$, and $s$ quarks and is electrically
neutral.  In the absence of electrons it seems difficult to
build a gap of high voltage between the CFL strange-matter core and
the outer crust.  However, if these  stars are in the 2SC phase,
they do 
contain electrons so that the formation of an electric dipole layer
at the surface of a superconducting 2SC strange quark matter core is
not a problem at all. 

Moreover,  in Ref.~\cite{Usov04} it has been shown that thin layers at the
surface of CFL strange-matter are no longer electrically neutral as
in the bulk because of surface effects \cite{Madsen01}. That is, in
the surface of CFL strange-matter the number of massive quarks is
suppressed relative to the number of massless quarks at fixed Fermi
momentum. This leads to a net increase in the total electrical charge
at the surface of CFL strange-matter. Therefore, we will
assume that  the Coulombic repulsion
mechanism for the halting of strange-matter core growth remains valid even for CFL 
or 2SC strange dwarfs.

\section{Results and Discussion}
\begin{figure}
\includegraphics[width=3.5in]{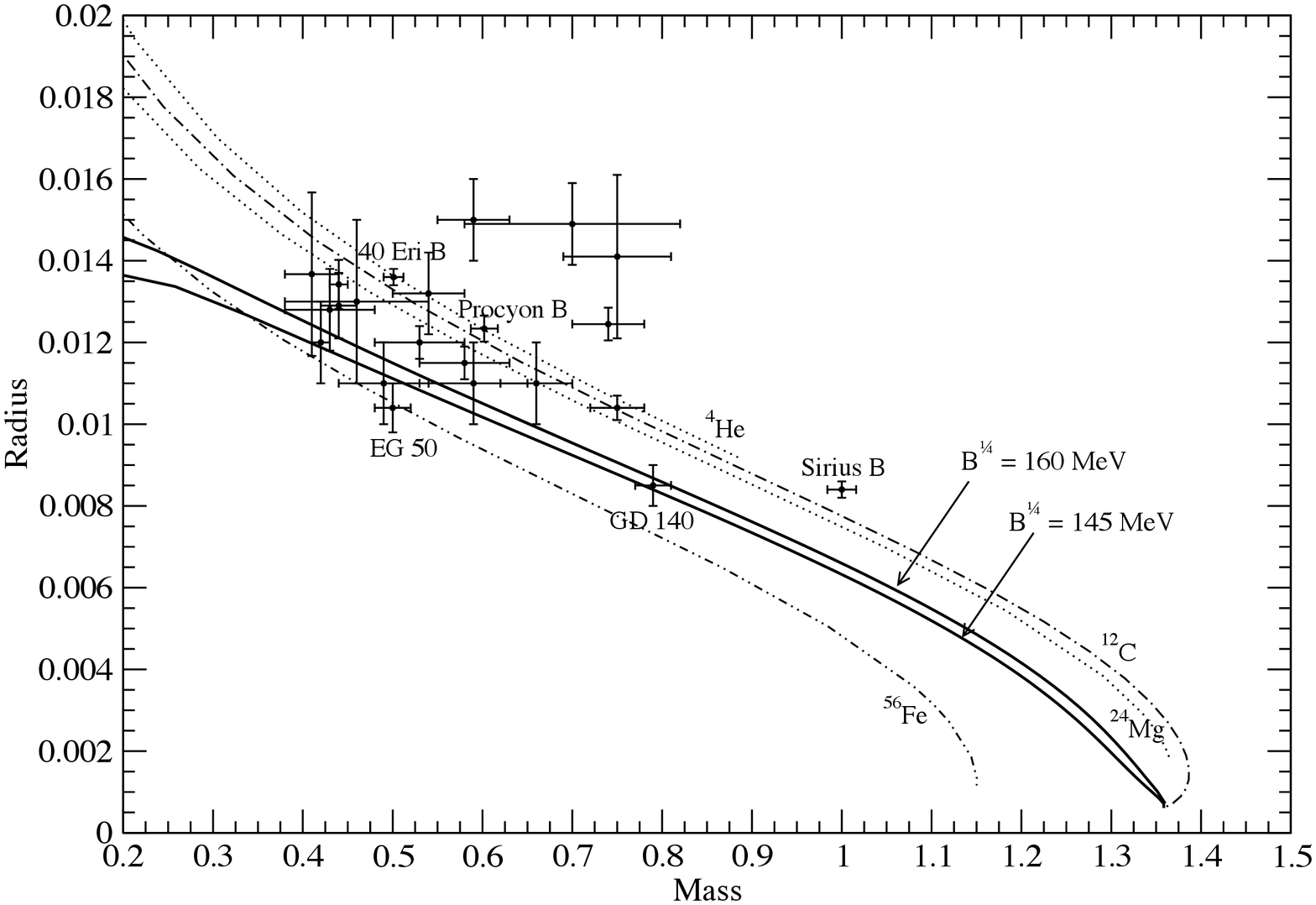} 
\caption[]{
Comparison of the theoretical mass-radius relationships for strange dwarfs (solid curves)
and normal white dwarfs with the data of \cite{Provencal98,Provencal02}.
}
\label{figmrsd}
\end{figure}

The mass-radius relation for strange-matter white dwarfs is shown in
Figure \ref{figmrsd} and compared with that derived from the 
normal-matter white-dwarf equation of state with various composition. The curves for strange
dwarfs agree  surprisingly well with several of the  best determined
data points.  Of greatest importance are {\it G238-44}, {\it EG 50}
and  {\it GD 140}. As mentioned earlier, these stars have well
determined masses, and they lie very close to the strange dwarf mass
radius relation. As a quantitative measure of the improvement to the
fit to the distribution.  We have evaluated the effects on the total
$\chi^2$ by associating each star with one curve or the other. The
stars which minimize $\chi^2$ by association with the strange-dwarf
mass-radius relation are identified as candidates in Table 1. The
total $\chi^2$ if all stars are associated with a normal carbon
white-dwarf EOS is 119 corresponding to a reduced $\chi^2_r = 5.2$
for 23 degrees of freedom. If we allow an iron-core population with
members chosen to minimize $\chi^2$ we have is 105 ($\chi^2_r = 4.4$
with 24 degrees of freedom). This is to be compared with a value of
78 ($\chi^2_r = 3.2$) when  eight stars are identified with a
strange-matter EOS instead. Hence, even allowing for the fact that
an extra degree of freedom has been introduced by assigning
membership in one population or the other, we have a $\Delta \chi^2
= 41~ (\sim 6\sigma)$ preference for the presence of a
strange-dwarf population over a normal EOS versus only a $\Delta
\chi^2 = 14~ (\sim 4\sigma)$ improvement with an iron-core
population.

\subsection{Observational tests}
Having identified these candidate stars as possible strange dwarfs
one would like to speculate on other observations which might be
used to discriminate between  a normal white dwarf and a strange
star. There are two possibilities which immediately come to mind.
Both of them are based upon astroseismology.  One could be the
filtering of those pulsation modes which are most sensitive to the
transition region between normal and strange matter.  The other
might be  the effects of strange-dwarf cooling rates on the observed
pulsations.

Regarding astroseismology, nonradial $g$-mode pulsations of white
dwarfs are observed to occur in three different phases of their
evolution (cf.~\cite{Brown94}). At temperatures above
80,000 K some hydrogen-deficient DOV white dwarfs and some planetary
nebulae are observed to pulsate. Also at cooler temperatures
$T_{eff} \approx 16,000$ to $25000$  K \cite{Winget83a,Winget83b}, 
some helium dominated DBV stars are
observed to pulsate.  Since almost all of the stars in the present
sample are DA white dwarfs with hydrogen-dominated spectra, it is
particularly noteworthy that a narrow strip exists
 \cite{Greenstein82,Bergeron95} for $T_{eff} \approx 12,000$ K in which DAV
(or $ZZ$ $Ceti$) stars are observed to pulsate. Unfortunately,
however, only one star ($G226$-$29$) in this sample falls within the
DAV pulsation strip and this star is not a candidate for a strange
matter core.  The pulsations of this star have been reanalyzed
\cite{Kepler00} using the {\it HST}.  Its inferred mass and
radius are consistent with a normal white dwarf and there is nothing
peculiar in the pulsation spectrum. On the other hand, we point out
that two of the strange-dwarf candidates, {\it G181-B5B} and {\it GD
279} are tantalizingly close (within uncertainty) to the DAV
instability temperature and may perhaps warrant further study to
search for peculiar pulsations.

If a pulsating strange-dwarf candidate were ever to be found, one
would expect its properties to be similar to a normal DAV with some
slight changes.  It would be similar because the bulk of the volume
of the star is still dominated by a normal matter crust as evident
in Figure \ref{wdprofile}.  On the other hand the development of a
steep abundance gradient in the normal matter for the inner
$\approx$1\% of the star followed by the sharp discontinuity at the
strange-matter core could have some observational consequences. One
might propose a test along the following lines.

For  idealized $g$-modes, the pulsations are approximately evenly spaced
with periods given by \cite{Kawaler94},
\begin{equation}
\Pi_{n l} = {\Pi_0 \over \sqrt{l(l+1)}} (n + \epsilon)~~,
\label{gmodes}
\end{equation}
with $\Pi_0$ a constant determined by the internal structure.
\begin{equation}
\Pi_0 = (2 \pi)^2 \biggl[ \int{N \over r}dr \biggr]^{-1} ~~,
\label{pi0}
\end{equation}
where $N$ is the Brunt V\"ais\"al\"a (buoyancy) frequency. The
simplest effect for a more compact star then is that the radial
integral in Equation (\ref{pi0}) changes.  On the one hand, the
integral extends over a smaller radius, while the
larger densities would imply a larger buoyancy frequency.
Anticipating that the latter effect is dominant, one would expect
shorter period oscillations for the more compact dwarfs.

Equation (\ref{gmodes}), however, only holds for stars with
homogeneous composition. The steep density gradient as one
approaches the strange-matter core will produce a rapid change in
the buoyancy frequency.  This will lead to mode trapping.  That is,
modes that have nodes in the region where the density is changing
rapidly will have the amplitude of their eigenfunctions
significantly reduced.  This small amplitude in the interior implies
that less energy is required to excite and maintain such modes. The
kinetic energy of a nonradial mode can be written \cite{Kawaler94}:
\begin{equation}
E_K \propto {\sigma^2 \over s} \int_o^R \rho r^2 dr
\bigl[ \xi_r^2 + l(l+1) \xi^2_h\bigr]~~,
\end{equation}
where $\xi_r$ and $\xi_h$ refer to the perturbation displacement in
the radial and horizontal directions, respectively.  The presence of
the density $\rho$ in the integral, however, implies that small
oscillations within the high-density core can contribute
significantly to the kinetic energy.

Thus, the smallest kinetic energies will be associated with trapped
nodes which have diminished interior amplitudes, but also with lower
modes with less displacement in the high-density core. This is
important since the growth rate for the amplitude of the mode scales
as $1/E_K$. Hence, these particular trapped modes will be the most
easily excited in the spectrum. Indeed, this filtering effect which
enhances the excitation of specific modes is well understood
\cite{Winget81} in normal DAV white dwarfs  where
this effect arises from  the hydrogen/helium discontinuity. Here we
speculate that a similar and perhaps more dramatic effect may occur
from the crust-strange-matter discontinuity.

Following this line or reasoning, we suggest that the periods of the
trapped modes in the outer layers of the star should obey,
\begin{equation}
\Pi_i^2 = 4 \pi^2 \lambda_i^2 \biggl[ \biggl( 1 - {r_{core} \over R}\biggr)
l(l+1) {G M \over R^3} \biggr]^{-1}~~,
\end{equation}
where $\lambda_i$ are constants relating to the zeros of Bessel
functions and the index $i$ corresponds to the number of nodes
between the surface and the core transition region which occurs at a
radius $r_{core}$. We propose therefore, that in principle one could
use the identification of these trapped modes to find the radius and
mass of the inner transition to a  strange-matter core.

Regarding cooling rate, it is generally appreciated that strange
neutron stars would cool more rapidly than normal neutron stars due
to the loss of interior energy by neutrinos produced from the weak
decays to strange matter \cite{Alcock86}. We
speculate that a similar process may be in operation early in the
life of a hot strange dwarf.  In such stars, the core may still be
growing to its putative equilibrium size and might be identified as
a hot pulsating star with an anomalously large cooling rate. 
If the core is in the CFL phase, it  has a superfluid gap that may be
as large as $\sim 100$ MeV. Such large gaps render the quark matter
core almost thermally "inert" since the heat capacity and neutrino
emissivity are proportional to exp(-kT/Delta). The cooling behavior of
strange dwarfs with crusts would thus almost entirely be determined by
the strange dwarf's nuclear crust and they would appear cool very quickly. 
On the other hand, if the core is
in the 2SC phase, which has a much smaller gap  ($\sim$ 
keV's to MeV's), the rapidly cooling quark matter core should have a significant impact
on the apparent thermal evolution.  Either way, the stars would appear to have
an anomalously rapid cooling rate.

Indeed, cooling rates have been observed \cite{Winget83a,Winget83b} 
in hot DOV stars as they evolve from planetary
nebulae to the white-dwarf cooling phase.   Such stars are evolving
so rapidly that one can measure changes in the pulsation periods
which relate directly to the cooling rate of the star.  We thus
propose that the identification of a compact DOV white dwarf with an
anomalously high cooling rate may constitute an independent
confirmation of the development of a strange-matter core.

An opposite effect, however, is possible if a strange dwarf is a
member of an accreting binary.  The strange dwarf would respond to
the accretion of matter by converting some of the material in the
crust to strange matter. The released latent heat would raise the
interior temperature and therefore require a longer timescale to
radiate away energy than that of a normal-matter white dwarf.

\section{Conclusions}

The purpose of the present study has been to explore the possible
existence of a new population of compact white dwarfs and whether
such compact stars are consistent with an interpretation that they
contain strange-matter cores. At present the data are too sparse and
uncertain to conclusively determine whether or not any of this
sample of DA  white dwarfs are strange dwarfs. Nevertheless, we have
shown that the deduced masses and radii are at least marginally
consistent with an interpretation that some stars in the sample
contain strange-matter cores.  Although this seems exotic, we have
argued that the alternative interpretation of iron-core white dwarfs
is difficult to achieve from a stellar evolution standpoint and does
not fit the observed compact population as well with a single curve.

Clearly, more data on the masses and radii for white dwarfs would be
of immense help in determining whether there exists a population of
compact white dwarfs.   Of particular interest would be the
discovery of a strange-dwarf candidate in the DAV (or DOV) pulsation
mode.  One might be able to detect the presence (or formation) of a
high density strange-matter core from the associated mode filtering.

\acknowledgments
The author's wish to acknowledge useful input from
Tom Guthrie and Doran Race in various aspects of the
data analysis. Work supported  in part by the US Department of Energy under
Nuclear Theory grant DE-FG02-95ER40934.  Work of B.O. supported in part from
the National Science Foundation a Research Experience for Uundergraguates grant at the University of
Notre Dame. One of the authors (KO) wishes to acknowledge support
from the Japanese Society for the Promotion of Science and by NSF grant PHY 02-16783 for the Joint Institute for Nuclear Astrophysics (JINA).  The research of
F. W. is supported in part by the National Science Foundation under Grant
PHY-0457329, and by the Research Corporation.


\vfill\eject
\begin{table}
\caption{Adopted$^a$ White Dwarf Properties}
\begin{tabular}{lccc}
{Star}           & {M/M$_\odot$}  & R/R$_\odot$ & $T (K)$\\
\tableline
\\
Normal White Dwarfs \\
\tableline
Sirius B     & $1.0034 \pm 0.026$ & $0.00840 \pm 0.00025$   & $24700 \pm 300$ \\
G226-29      & $0.750 \pm 0.030$ & $0.01040 \pm 0.0003$   & $12000 \pm 300$ \\
G93-48       & $0.750 \pm 0.060$ & $0.01410 \pm 0.0020$   & $18300 \pm 300$ \\
CD -38 10980 & $0.740 \pm 0.040$ & $0.01245 \pm 0.0004$   & $24000 \pm 200$ \\
L268-92      & $0.700 \pm 0.120$ & $0.01490 \pm 0.0010$   & $11800 \pm 1000$ \\
Stein 2051B  & $0.660 \pm 0.040$ & $0.0110  \pm 0.0010$  & $7100  \pm  50$ \\
Procyon B    & $0.602 \pm 0.015$ & $0.01234 \pm 0.00032$  & $7740  \pm  50$ \\
Wolf 485 A   & $0.590 \pm 0.040$ & $0.01500 \pm 0.0010$   & $14100 \pm 400$ \\
L711-10      & $0.540 \pm 0.040$ & $0.01320 \pm 0.0010$   & $19900 \pm 400$ \\
L481-60      & $0.530 \pm 0.050$ & $0.01200 \pm 0.0040$   & $11300 \pm 300$ \\
40 Eri B     & $0.501 \pm 0.011$ & $0.01360 \pm 0.0002$   & $16700 \pm 300$ \\
G154-B5B     & $0.460 \pm 0.080$ & $0.01300 \pm 0.0020$   & $14000 \pm 400$ \\
Wolf 1346    & $0.440 \pm 0.010$ & $0.01342 \pm 0.0006$   & $20000 \pm 300$ \\
Feige 22     & $0.410 \pm 0.030$ & $0.01367 \pm 0.0020$   & $19100 \pm 400$ \\
\\
\tableline
\\
Candidate Compact Strange Dwarfs\\
\tableline
GD 140       & $0.790 \pm 0.020$ & $0.00854 \pm 0.0005$  & $21700 \pm 300$ \\
G156-64      & $0.590 \pm 0.060$ & $0.01100 \pm 0.0010$   & $7160 \pm 200$ \\
EG 21        & $0.580 \pm 0.050$ & $0.01150 \pm 0.0004$  & $16200 \pm 300$ \\
EG 50        & $0.500 \pm 0.020$ & $0.01040 \pm 0.0006$  & $21000 \pm 300$ \\
G181-B5B     & $0.500 \pm 0.050$ & $0.01100 \pm 0.0010$  & $13600 \pm 500$ \\
GD 279       & $0.440 \pm 0.020$ & $0.01290 \pm 0.0008$  & $13500 \pm 200$ \\
WD2007-303   & $0.440 \pm 0.050$ & $0.01280 \pm 0.0010$  & $15200 \pm 700$ \\
G238-44      & $0.420 \pm 0.010$ & $0.01200 \pm 0.0010$  & $20200 \pm 400$ \\
\tableline

\end{tabular}

\end{table}
{$^a$}{Data from Provencal et al.~(1998; 2002).}

\end{document}